\documentstyle[12pt,weird]{article}
\begin{document}
%
\input epsf
\def\readfigfile#1#2{\centerline{\epsfysize=#2 \epsffile {#1} }}
%
\newcommand{\alphap}{\alpha^\prime}
\newcommand{\blambda}{\Lambda}
\newcommand{\bigxi}{\Xi}
\newcommand{\bomega}{\Omega}
\title{BLACK HOLES IN STRING THEORY}

\author{J.C.~Breckenridge\footnote{e-mail: jake@hep.physics.mcgill.ca}} 

\affil{Department of Physics,
McGill University, Montr\'eal, Qu\'ebec   \\H3A 2T8, CANADA}

\beginabstract
The black hole solutions to Einstein's vacuum field equations are also
solutions to the equations of motion of the low energy limit of
superstring theory.  At the same time, string theory boasts a much
broader and richer collection of black hole solutions.  Fortunately,
string theories also possess a remarkable set of duality symmetries
relating states within and between different string theories.  
These duality symmetries can be exploited to construct new black hole
solutions from known solutions, giving us powerful tools with which to
explore the black hole solutions of string theory.  Here we 
introduce and demonstrate these techniques of solution generating.
\endabstract

\section{INTRODUCTION}

Black holes are extremely interesting objects predicted by
Einstein's general theory of 
relativity as the endpoint of gravitational collapse.  
It has been shown that these objects
possess a thermodynamic entropy \cite{Bekenstein}, 
for which one would ideally like to have a microscopic and statistical 
interpretation.  Classical general relativity offers
no clues as to what this interpretation might be.
String theory, however, is at present a strong candidate for
a theory of quantum gravity.  
It is logical, therefore, to study black holes in the context of string
theory, in the hope that some light may be shed on 
the physics of black holes, in particular their entropy,
and in doing so validate the theory of strings as a physical theory.

Some progress has in fact been made on this front.
The dualities of strings \cite{drama} require additional objects
to be added to the theory, one species of which are the $D$-branes
\cite{dbranes}.
These are objects extended in zero or more dimensions, and
have been recently used to compute, for the first time, the
statistical entropy of black holes.  
Thus it is clear that $D$-branes represent useful probes into string theory,
in addition to their role 
in filling duality multiplets, serving as carriers of Ramond-Ramond charge
\cite{rcharge}. 

However, the $D$-brane technology is as yet applicable only to
a small class of extremal, supersymmetric black holes
\cite{blackhole,blackholes, blackholess},
although the method has been applied to solutions perturbed
slightly away from the extremal and supersymmetric limit \cite{nonext}.
One must therefore
find black hole solutions for which the method applies. 
Here we will illustrate the methods by which black hole solutions
amenable to $D$-brane analysis, and which therefore serve as
a testing ground for the method, can be obtained.

In section 2 we give a brief introduction to string theory symmetries.
Section 3 then treats in detail the low energy effective actions
and the duality symmetries to be used here.  The procedure used to
construct an example solution is then given, followed by a few
remarks about the characteristics of the solution generated, a
charged, rotating black hole in five spacetime dimensions.

\section{STRING THEORY SYMMETRIES}

One of the most fascinating aspects of the theory of strings is
the number of symmetries \cite{Giveon}
which can relate different regimes of
a given string theory to each other, or
relate one string theory to another. Recently, these
have had a crucial role to play in 
understanding the structure of string theory.
They have also led to the development of techniques 
which can be used to construct  new
solutions of the low-energy superstring equations of motion.

\begin{figure}
\label{dualitynetfig}
\readfigfile{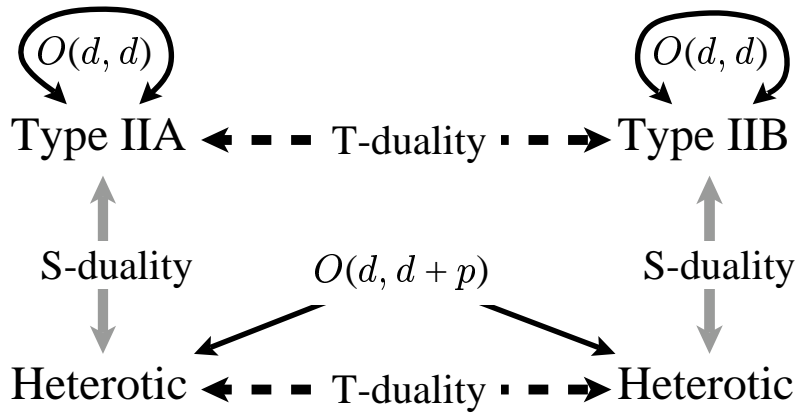}{2.2in}
\caption{An illustration of duality relations in string theory.}
\end{figure}

Here we divide these
symmetries, often known in the literature as
{\it dualities,\/} into three main groups:
\begin{itemize}
\item[$\bullet$] $O(d,d)$ symmetries are symmetries 
of the low-energy equations of motion which result from
independence of solutions of higher compact dimensions. In a sense we can
``rotate'' and/or ``boost'' a solution in such a way that it
becomes a different solution.  $O(d,d)$ keeps solutions within a given
string theory, rather than moving them between theories.
\item[$\bullet$] $T$-duality  symmetry inverts
the radius of compact dimensions,
i.e., $R \to 1/R$. Because a string
may wrap around a compact dimension, its spectrum of states 
is invariant under this change.  This feature has no analog 
in point particle theory.  $T$-duality sometimes moves a solution from
one theory to another, but not always.
\item[$\bullet$] $S$-duality symmetry relates weak- and strong-coupling 
regimes of a string theory, or different string theories.  
As we will see, under $S$-dualities the string coupling 
constant $g = e^\phi$ is inverted, $g \to 1/g$, thus leading to the 
interchange of strong- and weak-coupling physics.
\end{itemize}

In figure 1 we illustrate a {\it small\/} subset 
of the symmetries existing within and between string theories.   
For example, on the left we have an $S$-duality between the 
type IIA string and
the heterotic string, one of a set of string/string dualities.  
This duality occurs in six dimensions
between the heterotic string compactified from ten dimensions on the 
four-torus $T^4$, while the type IIA theory related to it is that
compactified on the Calabi-Yau manifold $K3$.  This duality,
with $O(d,d,{\bf R})$ symmetry, will be 
particularly useful here.

\section{SOLUTION GENERATING\label{sect.solgen}}

Let us note a few salient features of low--energy actions for the heterotic
and type IIA theories in six and five dimensions.  We work with
simplified versions of the six dimensional actions of the heterotic string
compactified on $T^4$ and the
type IIA string compactified on $K3$. 
All abelian gauge fields except one have been
set to zero, and all scalars resulting from compactification,
the moduli, are vanishing.  
The remaining gauge field is taken to be
an 
internal gauge field on the
heterotic side, and a Ramond--Ramond field,
associated with $D$-branes, on the Type II
side.

The low energy action of the heterotic string compactified on $T^4$ may be written \cite{senreview}:
\begin{equation}
S_{h} = \int\!\! d^6x \sqrt{-{\cal G}_6} 
e^{-2\phi^{(h)}_{6}} 
\left \{ R + 4 \big( \nabla \phi^{(h)}_6 \big)^2 
- {{1}\over{12}} (H^{(h)}_{6})^2
- {{1}\over{4}}{\cal F}_{6}^2              
\right \}
\label{six-un}
	\end{equation}
where ${\cal G}_6$ is the metric, $\phi^{(h)}_{6}$ is the dilaton,
$H^{(h)}_{6}$ is given by
\begin{equation}
H^{(h)}_{6\,\mu\nu\lambda}= \partial_\mu B^{(h)}_{6\,\nu\lambda}
- {{1}\over{2}}{\cal A}_{6\,\mu} {\cal F}_{6\,\nu\lambda} 
+ ({\rm cyclic})
\label{six-deux}
	\end{equation}
where $B^{(h)}_{6}$ is the antisymmetric tensor field and 
${\cal F}_{6} = d {\cal A}_6$ is the field strength of
the heterotic gauge field ${\cal A}_6$.

For the Type IIA theory, we write the action 
compactified on $K3$ as
\begin{eqnarray}
S_{IIA}  &=& \int\!\! d^6x 
\bigg[ \sqrt{-G_6} 
\bigg[ e^{-2\phi^{(a)}_{6}} 
\bigg( R + 4 \big( \nabla \phi^{(a)}_{6} \big)^2 
    - {{1}\over{12}}(H^{(a)}_{6})^2 
\bigg) \nonumber\\
&&\qquad \quad\mbox{} - {{1}\over{4}}F_{6}^2 \bigg] 
+ {{1}\over{16}} \epsilon^{\mu\nu\lambda\rho\alpha\beta}
B^{(a)}_{6\,\mu\nu} F_{6\,\lambda\rho} F_{6\,\alpha\beta}
\bigg]
\label{six-trois}
	\end{eqnarray}
where here $G_6$ and $\phi^{(a)}_{6}$ are the metric and dilaton,
while we have
\begin{equation}
H^{(a)}_{6\,\mu\nu\lambda} = \partial_\mu B^{(a)}_{6\,\nu\lambda} 
+ ({\rm cyclic})
\label{six-quatre}
	\end{equation}
as the field strength of the antisymmetric tensor field $B^{(a)}_{6}$.
These two actions are related by the string/string duality relation given 
as
\begin{eqnarray}
\phi^{(a)}_{6} &=& -\phi^{(h)}_6 \\
G_{6\mu \nu} &=& \e^{-2\phi^{(h)}_6} {\cal G}_{6 \mu \nu}\\
A_{6 \mu} &=& {\cal A}_ {6\mu}\\
H^{(a)}_{6\mu \nu \rho} &=& {1 \over 6} \epsilon_{ \mu \nu \rho \sigma
\kappa \delta}\, \e^{- 2\phi^{(h)}_6} H^{(h)\, \sigma \kappa \delta}_6,
\label{six-cinq}
	\end{eqnarray}
and the manifestation of strong-weak coupling exchange is then evident in
(5).
Since we also use the five-dimensional type IIA action 
compactified on $K3 \otimes S^1$,
we give here the standard Kaluza--Klein reduction on the circle $S^1$ with
coordinate labelling $y=x^5$,  
\begin{eqnarray}
ds_6^2 &=& {{G}}_{\mu\nu} dx^\mu dx^\nu 
+ e^{2\sigma} (dy + \blambda_{G\mu} dx^\mu)^2 \nonumber\\
\phi_6 &=& \phi + {{1}\over{2}}\sigma \nonumber\\
B_{6}  &=& {{1}\over{2}} [B_{\mu\nu} - {{1}\over{2}}
(\blambda_{G\mu} \blambda_{B\nu} 
- \blambda_{B\mu} \blambda_{G\nu})]dx^\mu \wedge dx^\nu
+ \blambda_{B\mu} dx^\mu \wedge dy
\label{six-six}
	\end{eqnarray}
where $\blambda_{G\mu}$ and $\blambda_{B\mu}$ are the gauge fields
coming from the compactification of the metric and antisymmetric tensor 
fields respectively.
The five dimensional type IIA action 
(in the sector with $A_y=0$) is expressed in the string frame 
as\footnote{We omit for simplicity the subscripts on the 
five-dimensional fields.}  
\begin{eqnarray}
S_{IIA} &=& \int\!\! d^5x 
\left[ \sqrt{-G} 
  \left[ e^{-2\phi^{(a)}} 
    \left({{R}} + 4 \big( \nabla \phi^{(a)} \big)^2 
    - {{1}\over{12}} (H^{(a)})^2 
    \right.
  \right. 
\right. \nonumber\\          &&\mbox{} 
\left. 
  \left. 
    \left. \qquad\qquad
    - (\partial_\mu \sigma)^2
    - {{1}\over{4}} e^{2\sigma} (\bigxi_{G})^2
    - {{1}\over{4}} e^{-2\sigma} (\bigxi_{B})^2 
    \right)
  \right. 
\right. \nonumber\\          &&\mbox{} 
\left. 
  \left. \qquad\qquad
  - {{1}\over{4}} e^{\sigma} F^2 
  \right] 
+ {{1}\over{8}}\epsilon^{\mu\nu\lambda\alpha\beta} 
\blambda_{B\mu} F_{\nu\lambda} F_{\alpha\beta} 
\right]
\label{six-sept}
	\end{eqnarray}
where $\bigxi_G = d \blambda_{G}$, $\bigxi_B = d \blambda_{B}$ and
\begin{eqnarray}
H^{(a)}_{\mu\nu\lambda} = \partial^{\vphantom{(a)}}_\mu B^{ (a)}_{\nu\lambda} 
- {{1}\over{2}} \blambda_{G\mu} \bigxi_{B\nu\lambda} 
- {{1}\over{2}} \blambda_{B\mu} \bigxi_{G\mu\nu} 
+ ({\rm cyclic}).
\label{six-huit}
	\end{eqnarray}
In five dimensions the Einstein frame metric is obtained from that
appearing in (\ref{six-sept}) through the transformation 
$g_{\mu\nu}=e^{-4\phi/3}{G}_{\mu\nu}$.

The solution of the five dimensional
theory with which we will begin is a five dimensional black hole 
spinning in a single plane. It is a solution of the 
vacuum Einstein equations, and is found in
\cite{Myers}.  We add a trivial flat compact dimension with coordinate 
$y$, and the six dimensional metric is then 
\begin{eqnarray}
ds_6^{2} 
&=& - dt^2 + (r^2+a^2)\sin^2\theta d\varphi^2
+ {\beta\over{\rho^2}}(dt + a\sin^2\theta d\varphi)^2 \nonumber\\
&&\mbox{} \qquad + {{\rho^2}\over{r^2 + a^2 - \beta}} dr^2
+ \rho^2 d\theta^2 + r^2 \cos^2\theta d\psi^2 + dy^2
\label{six-onze}
	\end{eqnarray}
where $\rho^2 = r^2 + a^2 \cos^2 \theta$, and $\beta$ and
$a$ are the mass and angular momentum parameters.
We are using spherical polar coordinates in five dimensions,
$t, r, \theta, \varphi, \psi$ with the additional coordinate $y$.
This black hole is a solution of the six dimensional
low energy type IIA string theory,  with only the metric excited. No gauge fields, antisymmetric tensor,
dilaton, or moduli fields are turned on.  From it, we will obtain 
a charged spinning black hole solution of the type IIA theory in five
dimensions, a spinning generalization of the 
solution in \cite{blackhole}.

\subsection{Generating techniques}

To generate the desired black hole solution
we use a series of transformations, namely $O(6,6,{\bf R})$
boosts involving the time $t$ and the circle
coordinate $y$, and string/string duality.  String/string
duality is implemented simply by computing the mapping 
given in equation (\ref{six-cinq}). For the $O(d,d,{\bf R})$
transformations, the procedure we
have implemented is that
outlined in \cite{HassanSen}.
Let us consider first the procedure for the heterotic
string with one non-zero gauge field, the extension to more than
one gauge field is straightforward.
One first forms, from the fields of the solution with which one begins, 
the linear combinations
\begin{equation}
{K_{\pm}}_{\mu \nu} = -B^{(h)}_{\mu \nu} - {\cal G}_{\mu \nu}
- {1 \over 4}{\cal A}_ \mu {\cal A}_ \nu \pm \eta_{\mu \nu}
\label{six-douze}
	\end{equation} 
from which the following matrix is constructed
\begin{eqnarray}
{\cal M} = 
\left(\begin{array}{ccc}K_{-}^t{\cal G}^{-1}K_{-} & K_{-}^t {\cal G}^{-1}K_{+} 
		& -K_-^t{\cal G}^{-1} {\cal A}\\
		K_{+}^t{\cal G}^{-1}K_{-} & K_{+}^t{\cal G}^{-1}K_{+}
		& -K_+^t{\cal G}^{-1} {\cal A} \\
		-{\cal A}^t{\cal G}^{-1} K_- & -{\cal A}^t
			{\cal G}^{-1} K_+ 
		& {\cal A}^t{\cal G}^{-1} {\cal A} \end{array}\right)
\label{six-treize}
	\end{eqnarray}
where the superscript $^t$ indicates the transpose.
At this point the effect of an $O(d,d,{\bf R})$ transformation on the
solution in question is contained in the relation
\begin{eqnarray}
{\cal M} \rightarrow \tilde{\cal M} = \bomega_d{\cal M}\bomega_d^t
\label{six-quatorze}
	\end{eqnarray}
where the transformation matrix $\bomega_d \in O(d,d,{\bf R})$.

After equation (16) has been computed,
there remains the question of extracting the new metric
and other fields from the
new matrix ${\tilde{\cal M}}$.  The way to do this is also found 
in \cite{HassanSen} and consists of forming the matrix
\begin{equation}
V = \left(\begin{array}{ccc} \eta/ 2 & - \eta/2 & 0\\
1/2 & 1/2 & 0 \\
0 & 0 & 1/\sqrt 2\end{array}\right)
\label{six-quinze}
	\end{equation}
which leaves $\tilde{\cal M}$ in a state looking like
\begin{equation}
{\tilde{\cal M}}^\prime = V\tilde{\cal M}V^t 
=\left(
\begin{array}{ccc}{{\cal G}^ \prime}^{-1} & -{{\cal G}^ \prime}^{-1}K^ \prime
& {{\cal G}^ \prime}^{-1} {{\cal A}}^\prime/ \sqrt{2} \\
-{K^ \prime}^t {{\cal G}^ \prime}^{-1} & {K^ \prime}^t{{\cal G}^ \prime}^{-1}K^ \prime 
& -{K^ \prime}^t {{\cal G}^ \prime}^{-1} {{\cal A}}^\prime/ \sqrt{2}\\
{{{\cal A}}^\prime}^t {{\cal G}^ \prime}^{-1}/ \sqrt{2} &
-{{{\cal A}}^\prime}^t {{\cal G}^ \prime}^{-1} K^\prime/ \sqrt{2}
& {{{\cal A}}^\prime}^t {{\cal G}^ \prime}^{-1} {\cal A}/ 2 \\
\end{array}\right)
\label{six-seize}
	\end{equation}
where $K^ \prime = ({K^ \prime}_{\pm}\mp \eta)$
and therefore the new metric, new $K^\prime$, and
gauge field $A$ can be extracted from the upper left,
upper center, and upper right parts respectively
of $\tilde{\cal M}^\prime$.  Then the antisymmetric tensor field
and dilaton are
\begin{equation}
{B^{(h)}}^ \prime =\, -{1 \over 2}(K^ \prime - {K^ \prime}^t), \qquad\qquad
\e^{2{\phi^{(h)}}^ \prime} =\, 
{ \det {\cal G}^ \prime \over \det {\cal G}}\e^{2 {\phi^{(h)}}}.
\label{six-dix-sept}
	\end{equation}

For the heterotic string the
group under which the equations of motion of the low energy
effective action are invariant is $O(d,d+p,{\bf R})$
where $d$ is the number
of coordinates in the full ten dimensional theory with respect to
which the solution is independent, and $p$ can be thought of as the number 
of gauge fields in the solution.  In the present case, therefore, 
on the heterotic side the group in question is $O(6,7,{\bf R})$.

What of the type IIA side of the story?  
$T$-duality, when applied
to a type II solution changes its chirality, altering 
its Ramond-Ramond field content.
Since $T$-duality is in the $O(d,d,{\bf Z})$
subgroup of $O(d,d,{\bf R})$, it is not possible to carry out the more
general $O(d,d,{\bf R})$ transformations on the RR sector.  
Therefore, one can apply this technique to the type II theories 
only when the Ramond-Ramond fields all vanish.  In this case
the formulae (\ref{six-douze}) through (19) apply when all
${\cal A}^{(1)\, \alpha} = 0$, giving $O(6,6,{\bf R})$ symmetry.

Let us now outline the
series of steps that will produce our new solution. 
We begin with
the metric (\ref{six-onze}) as a 
string-frame\footnote{Since the dilaton vanishes string frame and
Einstein frame metrics are identical.}
type IIA solution in six dimensions. 
We apply an $O(6,6)$ boost mixing the $(t,y)$ directions, 
following the five dimensional black hole construction 
of \cite{horsen}.
This causes the resulting six--dimensional 
solution to have no $G_{6\,y\mu}$ for $\mu<5$,
but induces a $B^{(a)}_{6 yt}$ and a $\phi^{(a)}_6$. 

The next step is to apply
string/string duality to create a heterotic solution from the 
type IIA solution.  This is done by applying the mapping (\ref{six-cinq}),
after which the new $B^{(h)}$ is computed up to gauge transformations
by integrating the field strength $H^{(h)}$. 
Using now the $O(6,7)$ symmetry we apply a second boost, 
mixing the time $t$ and the internal direction involving ${\cal A}_6$, 
with parameter $z$.  String/string duality is then applied a second
time to convert the heterotic solution back to a Type IIA solution, 
followed by the standard Kaluza--Klein reduction to five dimensions
as given in (\ref{six-six}), which in turn is followed by transformation
to the Einstein frame. 

The above boost parameters $x$ and $z$ are carefully chosen to 
satisfy $z=2x^2-1$ which reduces the five dimensional
dilaton to a constant.
The resulting configuration is a charged
spinning five dimensional black hole with constant dilaton.  
This solution is quite complicated and will not be written here,
but it simplifies substantially in the extremal limit.

\subsection{The extremal black hole}

The extremal limit of the black hole is computed by
taking the boost parameter $x$
off to infinity, and simultaneously the mass parameter 
$ \beta$ and the angular momentum parameter $a$ to zero
such that the quantities
\begin{equation}
\lim_{\stackrel{\scriptstyle x\to\infty}{\beta\to 0}}
 \beta\,x^2\equiv  \,\mu, \qquad\qquad
\lim_{\stackrel{\scriptstyle x\to\infty}{ a\to 0}}
a\,x\equiv \,\omega ,
\label{six-vingt-et-un}
	\end{equation}
remain finite, where $\beta,a$
are the quantities appearing in the metric (\ref{six-onze}).  
After doing a coordinate transformation to match with \cite{blackhole}, 
$r^2 \to r^2 + \mu$, we obtain for the extremal metric and
gauge fields
\begin{eqnarray}
ds_5^{2} 
&=& -{\left(1 - { \mu \over{r^2}} \right)^2} 
\left[dt - {{\mu\omega\sin^2\theta}\over{(r^2-\mu)}} d\varphi
+ {{\mu\omega\cos^2\theta} \over{(r^2-\mu)}} d\psi \right]^2 \nonumber\\
& &\mbox{} + \left( 1 - {\mu\over{r^2}} \right)^{-2}\! dr^2\!
+\! r^2 (d\theta^2 + \sin^2\theta d\varphi^2 + \cos^2\theta d\psi^2 ) \\
{A} &=& 
{\sqrt{2}\over\lambda}{{\mu}\over{r^2}}
\bigg ( dt +\omega\sin^2\theta d \varphi - \omega\cos^2\theta d \psi \bigg )\\
{\blambda_B} &=& 
{{\lambda^3}\over{\sqrt{2}}} {A}\\ 
e^{\sigma+2\phi^{(a)}/3} &=&\lambda^2 .
\label{six-vingt-deux}
	\end{eqnarray}

While the result of our solution generating procedure yields
zero dilaton and modulus, $\phi^{(a)} = 0 = \sigma$, we have shifted 
these scalars by a constant to those of (24),
which introduces the scaling of the gauge fields by $\lambda$
given above \cite{blackhole}. The above fields are the only ones 
excited in this black hole background.
Notice that when we take $\omega\to 0$, we recover the solution of
\cite{blackhole}.

\subsection{Properties of the solution}

The angular momenta in
the independent planes defined by $\varphi,\psi$ are
\begin{equation}
J_1 \equiv J_{\varphi} = +{\pi\over{4}} \mu \omega, \qquad\qquad
J_2 \equiv J_{\psi}    = -{\pi\over{4}} \mu \omega 
\label{six-vingt-trois}
	\end{equation}
and for the ADM mass we find
\begin{equation}
M_{ADM} = {{3\pi\mu}\over{4}},
\label{six-vingt-quatre}
	\end{equation}
while the charges under the form fields $ \bigxi$ and
$F$ are\footnote{The sphere $S^3$ is at infinity, so we
can ignore the effects of the Chern--Simons terms.}
\begin{eqnarray}
q_ \bigxi &\equiv& {{1}\over{4\pi^2}} \int_{S^3} 
{}^\star (e^{-2\sigma-4\phi^{(a)}/3}\, \bigxi_B) = \mu/\lambda^2 ,
\nonumber\\
q_ F &\equiv& {{1}\over{16\pi}} \int_{S^3}
{}^\star (e^{\sigma+2\phi^{(a)}/3} F) 
= -{{\pi}\over{2\sqrt{2}}}\mu\,\lambda .
\label{six-vingt-cinq}
	\end{eqnarray} 
Note that this black hole, although a solution of the low--energy
string theory equations, is not a solution of the Einstein--Maxwell
equations in five dimensions.  In the spinning configuration, the
magnetic dipole field combines with the electric monopole field so
that the Chern--Simons contributions to the equations of motion
are nontrivial.  

Let us now obtain the classical entropy of this extremal spinning black hole.
In the above coordinates, the horizon is at $r=r_0=\sqrt{\mu}$, and
its entropy is found to be ($|J_1|=|J_2|\equiv J$)
\begin{equation}
{\cal S} = {{1}\over{2}} \pi^2 \mu\sqrt{\mu - \omega^2}
       = 2\pi\sqrt{{{q_ \bigxi\,q_F^2}\over{2}}-J^2} 
\label{cinq-qqc}
	\end{equation}
where in the second line we have written the entropy
in terms of the physical charges and angular momenta.
Note that both of these expressions are independent of $\lambda$.
Counting of the entropy from a microscopic standpoint 
using $D$-brane methods reproduces exactly this result \cite{blackholess}.

This extremal rotating charged black hole
has a horizon with finite area, a feature not easy to find.
Ordinarily the addition of rotation (without energy)
to an extremal Reissner-Nordstrom black hole
destabilizes the horizon and yields a naked singularity. However
string theory stabilizes the horizon with the help of
a Chern-Simons coupling in the low-energy field theory.

\section{CONCLUSIONS}

We have demonstrated here the utility of the symmetries of 
low energy effective string theory in the construction of 
novel black hole solutions.  Thus these symmetries have a 
very practical application, allowing us in effect to solve
the equations of motion of string theory using purely
algebraic methods.  A demonstration of the power of these
symmetry methods was made by constructing a new class
of five-dimensional supersymmetric rotating black hole solutions.

One might say that the symmetries of string theory do triple duty.
First, they require the inclusion in string theory of additional
states beyond fundamental strings
called $D$-branes, which carry Ramond-Ramond charge.
These $D$-branes allow us to compute from a microscopic,
statistical standpoint the entropy of at least particular classes 
of black holes.  
Second, the symmetries give rise to methods of solution
generating by which new black hole solutions  may be constructed.
It is then possible to construct new black hole solutions which
fall into those classes to which the $D$-brane methods apply,
namely (near) supersymmetric black holes, 
allowing the methods to be thoroughly tested.
Third, these benefits of a more immediate and practical nature come
in addition to the role these symmetries play in our further
discovery  of the structure of the theory of strings.

\section*{ACKNOWLEDGEMENTS}
The author would like to thank R.C.~Myers, R.R.~Khuri and 
N.~Kaloper for helpful discussions and constructive criticism.  
The financial support
of NSERC of Canada and le Fonds FCAR du Qu\'ebec is also
appreciated.


\begin{thebibliography}{20}
\bibitem{Bekenstein}Bekenstein, J.D., 1973,
{\it  Phys. Rev.}\ {\bf D7}, 2333;\newref
Hawking, S.W., 1975 
{\it  Commmun. Math. Phys.}\ {\bf 43}, 199.
\bibitem{drama}Duff, M.J., 1996,
{\it  Int. J. Mod. Phys.}\ {\bf A11}, 5623;\newref
Schwarz, J.H., 1997,
{\it  Nucl. Phys. Proc. Suppl.}\ {\bf 55B} 1.
\bibitem{dbranes}Johnson, C.V., 1997,
{\it Nucl. Phys. Proc. Suppl.}\ {\bf 52A}, 326;\newref
Polchinski, J., 1996, 
{\it TASI Lectures on $D$-branes,\/} lectures at TASI-96,
Boulder, Colorado, e-print hep-th/9611050.
\bibitem{rcharge}Polchinski, J., 1995,
{\it Phys. Rev. Lett.}\ {\bf 75}, 4724
\bibitem{blackhole}Strominger, A. and Vafa, C., 1996,
{\it  Phys. Lett.}\ {\bf B379}, 99.
\bibitem{blackholes}
Horowitz, G.T. and Strominger, A., 1996, 
{\it  Phys. Rev. Lett.}\ {\bf 77}, 2368.
\bibitem{blackholess}
Breckenridge, J.C., Myers, R.C., Peet, A.W.
and Vafa, C., 1997, {\it  Phys. Lett.}\ {\bf B391}, 93.
\bibitem{nonext}Horowitz, G.T., Maldacena, J.M. and Strominger, A., 1996,
{\it  Phys. Lett.}\ {\bf B383}, 151;\newref
Breckenridge, J.C., Lowe, D.A., Myers, R.C.,
Peet, A.W., Strominger, A. and Vafa, C., 1996, 
{\it  Phys. Lett.}\ {\bf B381}, 423.
\bibitem{Giveon}Giveon, A., Porrati, M.
and Rabinovici, E., 1994, 
{\it  Phys. Rep.}\ {\bf 244}, 77.
\bibitem{senreview}Sen, A., 1994, 
{\it  Int. J. Mod. Phys.}\ {\bf A9}, 3707.
\bibitem{Myers}Myers, R.C. and Perry, M.J., 1986,
{\it Ann. Phys.}\ {\bf 12}, 304.
\bibitem{HassanSen}Hassan, S.F. and Sen, A., 1992,
{\it Nucl. Phys.}\ {\bf B375}, 103.
\bibitem{horsen}Horowitz, G.T. and Sen, A., 1996,
{\it  Phys. Rev.}\ {\bf D53}, 808.
\end{thebibliography}
\end{document}